\documentclass[journal, doublecolumn, 10pt]{IEEEtran}
\usepackage{epsfig,latexsym}
\usepackage{float}
\usepackage{indentfirst}
\usepackage{amsmath}
\usepackage{amssymb}
\usepackage{xcolor}
\usepackage{algorithmic,algorithm}

\usepackage{times}
\usepackage{subfigure}
\usepackage{psfrag}
\usepackage{amssymb}
\usepackage{cite}
\usepackage{fancyhdr}
\usepackage{color}
\begin{document}%
\title{ {\huge  Sum-Rate Maximization for NOMA-Assisted Pinching-Antenna Systems}}
\author{Ziwu Zhou,~Zheng Yang,~\IEEEmembership{Member,~IEEE}, Gaojie Chen,~\IEEEmembership{Senior Member,~IEEE},~Zhiguo, Ding,~\IEEEmembership{Fellow,~IEEE}

\thanks{ Z. Zhou and Z. Yang are with the Fujian Provincial Engineering Technology
 Research Center of Photoelectric Sensing Application, Fujian Normal Univer
sity, Fuzhou 350117, China (e-mail: zzzfjnu@163.com; zyfjnu@163.com).}
\thanks{ Gaojie Chen is with the School of Flexible Electronics (SoFE) and State Key
Laboratory of Optoelectronic Materials and Technologies (OEMT), Sun Yat-sen
University, Shenzhen 518107, China (e-mail: gaojie.chen@ieee.org).}
\thanks{ Z. Ding is with the Department of Computer and Information Engineering,
 Khalifa University, Abu Dhabi, United Arab Emirates, and also with the De
partment of Electronic and Electrical Engineering, The University of Manch
ester, M1 9BB Manchester, U.K. (e-mail: zhiguo.ding@manchester.ac.uk).}
}

\maketitle

\begin{abstract}
In this letter, we investigate a non-orthogonal multiple access (NOMA) assisted downlink pinching-antenna system. Leveraging the ability of pinching antennas to flexibly adjust users' wireless channel conditions, we formulate an optimization problem to maximize the sum rate by optimizing both the users' power allocation coefficients and the positions of pinching antennas. The optimal power allocation coefficients are obtained in closed-form by using the Karush-Kuhn-Tucker (KKT) conditions. The optimization problem of pinching antenna placements is more challenging than the power allocation problem, and is solved by a bisection-based search algorithm. In particular, the algorithm first optimizes the antenna placements to create favorable channel disparities between users, followed by fine-tuning the antenna positions to ensure the phase alignment for users, thus maximizing the sum rate. Simulation results demonstrate that, compared to conventional-antenna systems, pinching antennas can significantly enhance the sum rate in NOMA scenarios, and the proposed bisection-based search algorithm can achieve a sum rate nearly equivalent to that of an exhaustive search.
\end{abstract}

\vspace{-0.5em}

\vspace{-0.7em}
\section{Introduction}
In the sixth-generation (6G) wireless communication, the antenna technology is evolving toward greater flexibility and adaptability to accommodate the demands of high data rates, diverse deployment scenarios, and highly dynamic environments\cite{tataria20216g}, leading to the so-called flexible antennas. Fluid-antenna and movable-antenna systems are two typical examples of flexible antennas, offering enhanced adaptability by adjusting an antenna array's shape and position to effectively combat the small-scale fading\cite{wong2020fluid}\cite{zhu2024modeling}. However, the key limitation of fluid-antenna and movable-antenna systems is that their adjustments, which are typically at the wavelength scale, are often insufficient to compensate large-scale path loss. Pinching-antenna systems, employing novel dielectric waveguides, can radiate signals at any position along a waveguide, creating tailored local communication areas around the radiation points\cite{suzuki2022pinching}. This enables users to benefit from strong line-of-sight (LoS) links, significantly enhancing channel controllability\cite{ding2024flexible}. In \cite{xu2025rate}, the authors considered the data rate maximization for downlink pinching-antenna systems with a single user, and proposed a two-step positioning approach for pinching antennas to reduce large-scale path loss and achieve small-scale phase alignment. In \cite{wang2024antenna}, the authors investigated a NOMA-assisted downlink pinching-antenna system, focusing on the optimal activation of pinching antennas along a dielectric waveguide to maximize system throughput. However, the positions of the pinching antennas were fixed in  \cite{wang2024antenna}, and the system is to select the required antennas through activation rather than determining the optimal deployment locations to maximize system performance.

In this work, we focus on addressing the challenge of maximizing the sum rate in a NOMA-assisted pinching-antenna system by leveraging the channel reconfiguration capability of pinching antennas. Since the proposed sum rate maximization problem is non-convex, alternative optimization is employed to optimize users' power allocation coefficients and the positions of pinching antennas separately. Specifically, by first fixing the positions of pinching antennas, the optimal power allocation coefficients are obtained closed-form by utilizing the Karush-Kuhn-Tucker (KKT) conditions. Based on the obtained power allocation coefficients, a bisection-based search algorithm is employed to find the optimal pinching antenna positions, where these positions simultaneously affect both the large-scale fading for the users and the phase alignment caused by in-waveguide and free-space propagation. This algorithm strategically determines the placement of pinching antennas, thereby constructing favorable channel disparities between the users and enhancing the sum rate. Simulation results demonstrate the advantage of pinching antennas assisted NOMA over conventional antennas assisted NOMA, with the sum rate of the proposed bisection-based search algorithm closely matching that of the exhaustive search method. 

 \vspace{-0.7em}
\section{System Model}\label{model}

\begin{figure}[t]
\begin{center}
\includegraphics[width=2.8in, height=1.3in]{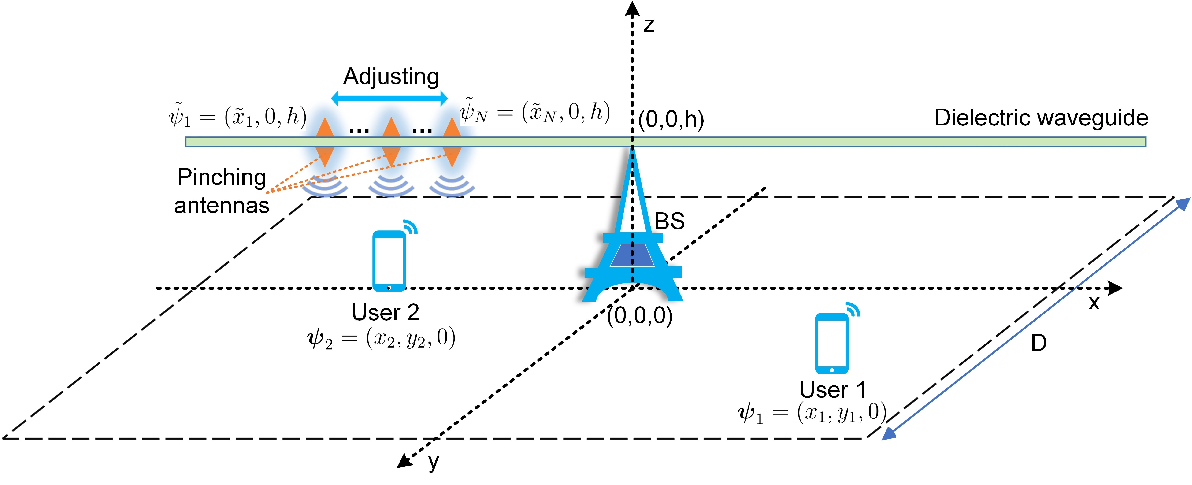}
\end{center}
\vspace*{-1.5em}
\caption{An illustration of the considered downlink NOMA-assisted pinching-antenna systems.}\label{system}
\vspace{-1em}
\end{figure}

As show in Fig. \ref{system}, we consider a downlink NOMA-assisted pinching-antenna system. Without loss of generality, we assume that a multi-antenna base station (BS) with a height of $h$ is located at the center of a square region, with its side length denoted by $D$, serving two single-antenna users simultaneously. The BS is equipped with a waveguide and $N$ pinching antennas. Specifically, the position of the $n$-th antenna is denoted by $\tilde{\psi}_{n}^\mathrm{P} = (\tilde{x}_{n}^\mathrm{P}, 0, h)$, where $\tilde{x}_{n}^\mathrm{P} \in [-D/2, D/2]$ and $n \in \mathcal{N} = \{1, \ldots, N\}$. To avoid antenna coupling, the spatial separation between any two antennas is set to be no less than $\Delta$. The two users are randomly distributed within the square region, and their coordinates are denoted by $\boldsymbol \phi_m = (x_m, y_m, 0), m\in \{1,2\}.$

\vspace{-0.2em}

By using the spherical wave channel model, the channel between the pinching antennas and User $m$ can be expressed as:
\begin{align}\label{pinching channel m}
\mathbf{h}_{m}^\mathrm{P} = \left[ \frac{\sqrt{\eta} e^{-j \frac{2 \pi}{\lambda} \left| \boldsymbol \phi_m - \tilde{\psi}_{1}^\mathrm{P} \right|}}{\left| \boldsymbol \phi_m - \tilde{\psi}_{1}^\mathrm{P} \right|}, \cdots, \frac{\sqrt{\eta} e^{-j \frac{2 \pi}{\lambda} \left| \boldsymbol \phi_m - \tilde{\psi}_{N}^\mathrm{P} \right|}}{\left| \boldsymbol \phi_m - \tilde{\psi}_{N}^\mathrm{P} \right|} \right]^\mathrm{T},
\end{align}
where $m\in \{1,2\}$, $\eta = \frac{c^2}{16 \pi^2 f_c^2}$, $c$ is the speed of light, $f_c$ is the carrier of frequency, and $\lambda$ is the wavelength in free space. Since all pinching antennas share the same waveguide with a single radio frequency (RF) chain, the signals intended for different users must be superimposed before transmission. According to the downlink NOMA principles, the superimposed signal transmitted through the waveguide can be expressed as:
\begin{align}\label{signal transmitted}
s = \sqrt{\alpha_1} s_1 + \sqrt{\alpha_2} s_2,
\end{align}
where $s_1$ and $s_2$ represent the signal intended for User $1$ and User $2$, respectively, with the power allocation coefficients denoted by $\alpha_1$ and $\alpha_2$, satisfying $\alpha_1 + \alpha_2 = 1$.

It is important to note that the propagation of signals along the waveguide introduces additional phase shifts. Consequently, the signal vector radiated from the pinching antennas at each radiation point can be expressed as:
\begin{align}\label{signal radiated}
\mathbf{x} = \sqrt{\frac{P_t}{N}}\left[e^{-j\theta_1}, \cdots, e^{-j\theta_N}\right]^\mathrm{T} s,
\end{align}
where $P_t$ denotes the transmit power of BS and $
\theta_n = 2\pi \frac{\left|\psi_0^\mathrm{P}-\tilde{\psi}_{n}^\mathrm{P}\right|}{\lambda_g}
$ accounts for the phase shift induced by signal propagation along the waveguide. $\psi_0^\mathrm{P}$ represents the position of the waveguide feed point, and  $\lambda_g = \frac{\lambda}{n_\text{eff}}$ denotes the guided wavelength, $n_\text{eff}$ denotes the effective refractive index of the dielectric waveguide. Note that $\frac{P_t}{N}$ represents the ideal transmission power for each pinching antenna, indicating that the total transmission power $P_t$  from the BS is uniformly distributed across the $N$ radiation points\cite{ding2024flexible}.

Based on \eqref{pinching channel m} and  \eqref{signal radiated}, the signal received by User $m$ can be expressed as:
\begin{align}\label{signal received}
y_{m}^\mathrm{P} = \left(\mathbf{h}_m^\mathrm{P}\right)^\mathrm{H} \mathbf{x} + w_m, m\in \{1,2\},
\end{align}
where $w_m$ denotes the additive white Gaussian noise (AWGN) with zero mean and variance $\sigma^2$. By substituting \eqref{pinching channel m} and \eqref{signal radiated} into \eqref{signal received}, the received signal $y_{m}^\mathrm{P}$ can be further rewritten as:
\begin{align}\label{signal received2}
y_{m}^\mathrm{P} = \sqrt{\frac{P_t}{N}} \mathbf{g}_{m} \left(\sqrt{\alpha_{1}} s_1 + \sqrt{\alpha_{2}} s_2\right) + w_{m}, m \in \{1,2\},
\end{align}
where
\begin{align}\label{g_m}
\mathbf{g}_{m} \triangleq \sum_{n=1}^{N} \frac{\sqrt{\eta} e^{j2\pi \left(\frac{\left|\boldsymbol \phi_m-\tilde{\psi}_{n}^\mathrm{P}\right|}{\lambda}-\frac{\left|\psi_{0}^\mathrm{P}-\tilde{\psi}_{n}^\mathrm{P}\right|}{\lambda_{g}}\right)}}{\left| \boldsymbol \phi_m-\tilde{\psi}_{n}^\mathrm{P}\right|}.
\end{align}

It is worth noting that in conventional-antenna systems, the antennas are fixed at the BS. The user with a stronger LoS link to the BS is typically viewed as the strong user, while the user with a weaker LoS link is viewed as the weak user. Nevertheless, in pinching-antenna systems, pinching antennas can emit signals from any position along the waveguide, allowing for flexible reconfiguration of the wireless channel conditions for users. For a single user, to maximize the user's rate, each pinching antenna is typically placed at the position along the waveguide that is closest to the user\cite{xu2025rate}. However, in this paper, two users are considered within the NOMA scheme, where the placement of pinching antennas must realize sufficient channel differentiation between users\cite{yang2025pinching}, thereby maximizing the sum rate while meeting the quality of service (QoS) requirements for each user. In an LoS propagation environment, a user positioned closer to the waveguide can take advantage of the dynamic controllability of pinching-antenna systems to achieve lower path loss and enhanced channel gains. In Fig. \ref{system}, since User $2$ is closer to the waveguide than User $1$, User $2$ can be identified as a potential strong user, while User $1$ is considered a potential weak user. Furthermore, it is assumed that User $2$ requires a high data rate, while User $1$ needs to be served at a low data rate only, implying that the channel conditions must satisfy  $|\mathbf{g}_2|^2 \geq |\mathbf{g}_1|^2$. In this scenario, the locations of the pinching antennas must be carefully designed to provide a strong LoS for User $2$, while satisfying the rate requirements of User $1$.


According to NOMA, User $1$ is expected to directly decode its own signal while treating the message of User $2$ as co-channel interference. On the other hand, User $2$ should perform successive interference cancellation (SIC), where it first decodes User $1$’s signal, removes the interference, and then decodes its own signal. Therefore, based on \eqref{signal received2}, the achievable data rate of User $1$ decoding its own signal can be expressed as:
\begin{align}\label{R1_P RATE}
R_1^{\mathrm{P}} = \log_2\left( 1 + \frac{\alpha_1 \frac{P_t}{N} |\mathbf{g}_1|^2}{\alpha_2 \frac{P_t}{N} |\mathbf{g}_1|^2 + \sigma^2} \right).
\end{align}

For User $2$, SIC is performed to decode the signal from User $1$, with the corresponding achievable rate expressed as:
\begin{align}\label{success SIC}
R_{2 \to 1}^{\mathrm{P}} = \log_2\left( 1 + \frac{\alpha_1 \frac{P_t}{N} |\mathbf{g}_2|^2}{\alpha_2 \frac{P_t}{N} |\mathbf{g}_2|^2 + \sigma^2} \right).
\end{align}

If User 2 can successfully perform SIC, i.e.,
$R_{2 \to 1}^{\mathrm{P}} \geq \tilde{R}_1^{\text{min}}$, where $\tilde{R}_1^{\text{min}}$ is the target data rate of User $1$, the signal of User $1$ can be correctly removed by User 2. Therefore, the achievable rate for User $2$ to decode its own message can be expressed as:
\begin{align}\label{R2_P RATE}
R_2^{\mathrm{P}} = \log_2\left( 1 + \frac{\alpha_2 \frac{P_t}{N} |\mathbf{g}_2|^2}{\sigma^2} \right).
\end{align}

\vspace{-0.7em}
\section{Problem Formulation and Proposal Solution}\label{solutions}
Our objective is to maximize the achievable sum rate, while ensuring that the minimum data rate requirements for each user are satisfied. As indicated by \eqref{R1_P RATE} and \eqref{R2_P RATE}, the achievable rates are influenced not only by the power allocation coefficients but also significantly by the locations of the pinching antennas, denoted by $ \mathbf{x}^{\mathrm{P}} = [\tilde{x}_1^{\mathrm{P}}, \ldots, \tilde{x}_N^{\mathrm{P}}]$. Therefore, the problem of maximizing the sum rate for NOMA-assisted pinching-antenna systems can be formulated as:
\begin{subequations} \label{opti.eq1s}
\begin{align}
\max_{\mathbf{x}^{\mathrm{P}}, {\alpha_1}, {\alpha_2}} \quad & R_1^{\mathrm{P}} + R_2^{\mathrm{P}}\\
\text{s.t.} \quad & |\tilde{x}_{n}^{\mathrm{P}} - \tilde{x}_{n-1}^{\mathrm{P}}| \geq \Delta, \ \forall n \in \mathcal{N},\label{antenna gap}\\
& R_1^{\mathrm{P}} \geq \tilde{R}_1^{\text{min}},\label{R1 min}\\
& R_2^{\mathrm{P}} \geq \tilde{R}_2^{\text{min}},\label{R2 min}\\
& R_{2 \to 1}^{\mathrm{P}} \geq \tilde{R}_1^{\text{min}}, \label{SIC sucess}\\
& \alpha_1 > \alpha_2 > 0,\label{SIC}\\
& \alpha_1 + \alpha_2 = 1,\label{pwoer allocation}\\
& |\mathbf{g}_2|^2 \geq |\mathbf{g}_1|^2,\label{channel requirement}
\end{align}
\end{subequations}
where $\tilde{R}_2^{\text{min}}$ denotes the target data rate of User $2$. Constraint \eqref{antenna gap} prevents potential inter-channel coupling caused by the physical spacing between antennas. Constraints \eqref{R1 min} and \eqref{R2 min} represent minimum target data rate requirements for the two users, respectively. Constraint \eqref{SIC sucess} ensures that the SIC can be correctly performed by User $2$. Constraints \eqref{SIC} and \eqref{pwoer allocation} reflect the   feasible power allocation coefficients for the users. Finally, constraint \eqref{channel requirement} represents the channel conditions for the SIC decoding order.

Based on \eqref{R1_P RATE}, \eqref{R2_P RATE}, \eqref{SIC} and \eqref{pwoer allocation}, the original problem \eqref{opti.eq1s} can be further simplified and expressed as:
\begin{subequations} \label{subopti.eq1s3}
\begin{align}
\max_{ {\alpha_2}, \mathbf{x}^{\mathrm{P}}} \quad & \log_2\left( 1 +  f\left(\alpha_2, \mathbf{x}^{\mathrm{P}}\right)\right)\label{objective function}\\
\text{s.t.} \quad & \frac{1}{2} > \alpha_2 > 0,\label{SIC1}\\
& \eqref{antenna gap}- \eqref{SIC sucess}, \eqref{channel requirement},
\end{align}
\end{subequations}
where $f\left(\alpha_2, \mathbf{x}^{\mathrm{P}}\right) \triangleq \alpha_2 \rho |\mathbf{g}_2|^2 + \frac{ \left ( 1 - \alpha_2 \right ) \left ( 1 + \alpha_2 \rho |\mathbf{g}_2|^2 \right ) \rho |\mathbf{g}_1|^2}{\alpha_2 \rho |\mathbf{g}_1|^2 + 1}$, and $\rho = \frac{P_t}{N \sigma^2}$.

Due to the coupling of variables in problem \eqref{subopti.eq1s3}, the  problem becomes non-convex, making it challenging to solve directly. To address the challenging problem, alternative optimization is applied to decompose problem \eqref{subopti.eq1s3} into two sub-problems which are solved alternately. Specifically, we first fix the positions of the pinching antennas to determine the optimal power allocation coefficients. Then, based on the optimized power allocation coefficients, we optimize the positions of the pinching antennas.
\vspace{-0.7em}
\subsection{Power Allocation  Coefficients Optimization}
\vspace{-0.2em}
Given the positions of pinching antennas  $\mathbf{x}^{\mathrm{P}}$, the problem \eqref{subopti.eq1s3} can be expressed as:

\begin{subequations} \label{subopti.eq1s2}
\begin{align}
\max_{ {\alpha_2}} \quad & f\left(\alpha_2\right)\label{objective function1}\\
\text{s.t.} \quad & \frac{ \left ( 1 - \alpha_2 \right ) \rho |\mathbf{g}_1|^2}{\alpha_2 \rho |\mathbf{g}_1|^2 + 1} \geq 2^{\tilde{R}_1^{\text{min}}} - 1,\label{R1 min2}\\
& \alpha_2 \rho |\mathbf{g}_2|^2 \geq 2^{\tilde{R}_2^{\text{min}}} - 1,\label{R2 min2}\\
& \frac{ \left ( 1 - \alpha_2 \right ) \rho |\mathbf{g}_2|^2}{\alpha_2 \rho |\mathbf{g}_2|^2 + 1} \geq 2^{\tilde{R}_1^{\text{min}}} - 1,\label{SIC sucess2}\\
& \frac{1}{2} > \alpha_2 > 0.\label{SIC1}
\end{align}
\end{subequations}

An important observation is that problem \eqref{subopti.eq1s2} can be reformulated as the maximization of a concave objective function $f\left(\alpha_2\right)$, where all the constraints are convex. Consequently, the corresponding Lagrangian function $\mathcal{L} \left( \alpha_2, \lambda_1, \lambda_2, \lambda_3 \right)$ can be expressed as:
\begin{align}
 \mathcal{L} \left( \alpha_2, \lambda_1, \lambda_2, \lambda_3 \right) &= f \left(\alpha_2 \right) + \lambda_1 \left( \alpha_2 - \frac{ \rho |\mathbf{g}_1|^2 + 1 - 2^{\tilde{R}_1^{\text{min}}}}{\rho |\mathbf{g}_1|^2  2^{\tilde{R}_1^{\text{min}}}}\right) \nonumber \\
 & + \lambda_2 \left( \alpha_2 - \frac{ \rho |\mathbf{g}_2|^2 + 1 - 2^{\tilde{R}_1^{\text{min}}}}{\rho |\mathbf{g}_2|^2  2^{\tilde{R}_1^{\text{min}}}}\right) \nonumber \\
 & + \lambda_3 \left( 2^{\tilde{R}_1^{\text{min}}} - 1 - \alpha_2 \rho |\mathbf{g}_2|^2 \right) ,
\end{align}
where $\lambda_1, \lambda_2, \lambda_3 \geq 0$ are the Lagrange multipliers. By applying the KKT conditions and with some straightforward algebraic manipulations \cite{boyd2004convex}, we can derive a closed-form solution $\tilde{\alpha}_2^* = \frac{ \rho |\mathbf{g}_1|^2 + 1 - 2^{\tilde{R}_1^{\text{min}}}}{\rho |\mathbf{g}_1|^2  2^{\tilde{R}_1^{\text{min}}}}.$

More significantly, for problem \eqref{subopti.eq1s3} to remain feasible, the power allocation  coefficient $\alpha_2$ must satisfy $\alpha_2 \in \left(0, \frac{1}{2}\right)$. If the solution $\tilde{\alpha}_2^*$ falls outside this range, it must be appropriately adjusted to ensure feasibility\cite{yang2017impact}. Therefore, the optimized  power allocation coefficient $\alpha_2^*$ can be expressed as:
\begin{align}
\alpha_2^* = \max \{0, \min \{ \tilde{\alpha}_2^*, 0.5\}\}.\label{opt_alpha_2}
\end{align}
\vspace{-0.7em}
\subsection{Pinching Antenna Positions Optimization}
\vspace{-0.2em}
 Based on the optimized power allocation coefficient $\alpha_2^*$ in \eqref{opt_alpha_2}, our goal is to optimize the positions of the pinching antennas $\mathbf{x}^{\mathrm{P}}$ to maximize the sum rate in problem \eqref{subopti.eq1s3}. According to \eqref{g_m}, the sum rate \eqref{objective function} monotonically increases with respect to $|\mathbf{g}_1|^2$ and $|\mathbf{g}_2|^2$, both of which are closely related to the placement of the pinching antennas $\mathbf{x}^{\mathrm{P}}$. Optimizing the pinching antennas' positions involves two steps: first, determining the optimal antenna positions on a macroscopic scale to maximize the sum rate under NOMA strategy; and second, fine-tuning the antenna positions on a wavelength scale to ensure precise phase alignment between the signal components propagated through free space and within the waveguide. Without loss of generality, we assume that the pinching antennas are arranged in a sequential order. Therefore, based on \eqref{opt_alpha_2}, the original problem \eqref{subopti.eq1s3} can be formulated as:
\begin{subequations} \label{subopti.eq1s1}
\begin{align}
\max_{ \mathbf{x}^{\mathrm{P}}} \quad & \alpha_2^* \rho |\mathbf{g}_2|^2 + \frac{ \left ( 1 - \alpha_2^* \right ) \left ( 1 + \alpha_2^* \rho |\mathbf{g}_2|^2 \right ) \rho |\mathbf{g}_1|^2}{\alpha_2^* \rho |\mathbf{g}_1|^2 + 1} \label{objective function2}\\ 
\text{s.t.} \quad &\tilde{x}_{n}^{\mathrm{P}} - \tilde{x}_{n-1}^{\mathrm{P}} \geq \Delta, \forall n\in \mathcal{N},\label{gap}\\
&\phi_{1, n} - \phi_{1, n - 1} = 2k\pi + \delta_1, \forall n \in \mathcal{N},\label{user1 aligned}\\
&\phi_{2, n} - \phi_{2, n - 1} = 2k\pi + \delta_2, \forall n \in \mathcal{N},\label{user2 aligned}\\
& \eqref{R1 min}, \eqref{R2 min}, \eqref{SIC sucess}, \eqref{channel requirement},
\end{align}
\end{subequations}
where $\phi_{m, n} = 2\pi \left(\frac{\left|\psi_{m}-\tilde{\psi}_{n}^\mathrm{P}\right|}
{\lambda}-\frac{\left|\psi_{0}^\mathrm{P}-\tilde{\psi}_{n}^\mathrm{P}\right|}{\lambda_{g}}\right), m\in \{1,2\}.$ Constraints \eqref{user1 aligned} and \eqref{user2 aligned} represent the phase alignment accuracy (modulo-2$\pi$) from the pinching antennas to User $1$ and User $2$, respectively, where $k$ is an arbitrary integer, and $\delta_1 $ and $\delta_2$ are predefined non-negative precision constants. Note that to maximize the sum rate and satisfy the QoS requirements in NOMA-assisted pinching-antenna systems, this is closely related to the proportion of resources allocated to User 2. Therefore, to achieve optimal phase alignment for User $2$, we assume that $\delta_2 \leq \delta_1$.

In problem \eqref{subopti.eq1s1}, the variation in the pinching antenna positions simultaneously affects the channel conditions of both users, and the objective function is non-convex, making direct analysis highly complex. To address this, we propose a bisection-based iterative search algorithm, with its detailed steps summarized in {\bf Algorithm} \ref{alg:1}.
In particularly, we leverage the strong LoS capability provided by the pinching antennas to differentiate User $1$ and User $2$, ensuring $|y_1| \geq |y_2|$. To maximize the achievable sum rate while meeting the QoS requirements of the users, resource allocation should be prioritized for User $2$ as much as possible. Therefore, the pinching antenna positions $\mathbf{x}^{\mathrm{P}}$ should be adjusted to approach the position of $x_2$. The details of {\bf Algorithm} \ref{alg:1} are presented as follows.

Initially, the left boundary of the search range for the antenna positions is set to be equal to the x-axis coordinate of User 2, i.e., $\gamma^{\text{left}} = x_2$, while the right boundary is set as the x-axis coordinate of User 1, i.e., $\gamma^{\text{right}} = x_1$. Without loss of generality, the initial position of the $(\frac{N + 1}{2})$-th pinching antenna is set at the midpoint of the x-axis coordinates of the two users (for odd $N$), while the remaining antennas are evenly distributed, ensuring the minimum physical spacing between them to avoid coupling. Subsequently, the positions of the left and right antennas are finely adjusted on the scale of wavelength $\tilde{\Delta}$ to achieve precise phase alignment for both users' signals, ensuring optimal propagation in free space and within the waveguide.

After fine-tuning the antenna positions, solving problem \eqref{subopti.eq1s2} yields the optimal power allocation coefficient $\alpha_2^*$ for the current antenna configuration. Based on \eqref{gap}, \eqref{user1 aligned}, and \eqref{user2 aligned}, the updated channel conditions are evaluated to verify whether the QoS requirements of the users are satisfied. If the QoS of User $1$ is met, the bisection method updates the right boundary $\gamma^{\text{right}}$ to the current position of the  $(\frac{N + 1}{2})$-th pinching antenna. Otherwise, the left boundary $\gamma^{\text{left}}$ is updated to the current position of the $(\frac{N + 1}{2})$-th pinching antenna. The process is repeated the left and right boundaries converge, at which point the algorithm terminates. The final results include the optimized power allocation coefficients $\alpha_2^*$, the optimized pinching antenna positions $\mathbf{x}^{\mathrm{P}^*}$, and the maximum achievable sum rate $R_1^{\mathrm{P}^*} + R_2^{\mathrm{P}^*}$. 

\begin{algorithm}
    \caption{The Iterative Algorithm for Solving Problem \eqref{subopti.eq1s1}}
    \label{alg:1}
    \begin{algorithmic}[1]
        \STATE \textbf{Initialize:} Initialize $\gamma^{\text{left}} = x_2$, $\gamma^{\text{right}} = x_1$. Set $\epsilon = 10^{-5}$, $\delta_1$, $\delta_2$, and $\tilde{\Delta}$.
        \STATE \textbf{while} $|\gamma^{\text{right}} - \gamma^{\text{left}}|> \epsilon$ \textbf{do}
        \STATE \quad $ \tilde{x}_{\frac{N + 1}{2}}^{\mathrm{P}} = \gamma^{\text{cur}} = \frac{1}{2}(\gamma^{\text{left}} + \gamma^{\text{right}})$.
        \STATE \quad$\tilde{x}_{n}^{\mathrm{P}} - \tilde{x}_{n-1}^{\mathrm{P}} = \Delta, \forall n \in \mathcal{N}$.
        \STATE \quad \textbf{for} $n = \frac{N + 1}{2} + 1 : N$ \textbf{do}
        \STATE \quad \quad Get $\theta_m = \text{mod}\{|\phi_{m, n} - \phi_{m, n - 1}|, 2\pi \}, m \in \{1,2\}$.
        \STATE \quad \quad \textbf{if} $\tilde{x}_{n}^{\mathrm{P}} - \tilde{x}_{n-1}^{\mathrm{P}} \geq \Delta$, $\theta_1 \leq \delta_1$, and  $\theta_2 \leq \delta_2$ \textbf{then}
        \STATE \quad \quad \textbf{break}
        \STATE \quad \quad \textbf{else}
        \STATE \quad \quad \quad $\tilde{x}_{n}^{\mathrm{P}} = \tilde{x}_{n}^{\mathrm{P}} + \tilde{\Delta}$.
        \STATE \quad \quad \textbf{end if}
        \STATE  \quad \textbf{end for}

        \STATE \quad \textbf{for} $n = \frac{N + 1}{2} - 1 : -1 : 1$ \textbf{do}
        \STATE \quad \quad Get $\theta_m = \text{mod}\{|\phi_{m, n} - \phi_{m, n + 1}|, 2\pi \}, m \in \{1,2\}$.
        \STATE \quad \quad \textbf{if} $\tilde{x}_{n}^{\mathrm{P}} - \tilde{x}_{n + 1}^{\mathrm{P}} \geq \Delta$, $\theta_1 \leq \delta_1$, and  $\theta_2 \leq \delta_2$ \textbf{then}
        \STATE \quad \quad \textbf{break}
        \STATE \quad \quad \textbf{else}
        \STATE \quad \quad \quad $\tilde{x}_{n}^{\mathrm{P}} = \tilde{x}_{n}^{\mathrm{P}} - \tilde{\Delta}$.
        \STATE \quad \quad \textbf{end if}
        \STATE \quad \textbf{end for}

        \STATE \quad Solve for $\alpha_2^*$ using \eqref{opt_alpha_2}.
        \STATE \quad \textbf{if} \eqref{R1 min}, \eqref{R2 min}, and \eqref{SIC sucess} \textbf{then}
        \STATE \quad \quad $\gamma^{\text{right}} = \gamma^{\text{cur}}$,
        \STATE \quad \textbf{else}
        \STATE \quad \quad $\gamma^{\text{left}} = \gamma^{\text{cur}}$.
        \STATE \quad \textbf{end if}
        \STATE \textbf{end while}
        \STATE \textbf{Output:} Obtain the optimized $\alpha_2^*$, $\mathbf{x}^{\mathrm{P}^*}$ and $R_1^{\mathrm{P}^*} + R_2^{\mathrm{P}^*}$.
    \end{algorithmic}
\end{algorithm}


\vspace{-0.7em}
\section{Simulation Results}\label{results}
In this section, we evaluate the sum rate of the pinching-antenna systems assisted by NOMA, as well as the effectiveness of the proposed algorithm. For the simulations, the parameters are configured as follows: the noise power is set to $-90$  dBm, $h = 3$ m, $f_c = 28$ GHz, $n_\text{eff}$ = 1.4, $\Delta = \frac{\lambda}{2}$, $N =3$, and $\tilde{R}_1^{\text{min}} = \tilde{R}_2^{\text{min}} = 0.5$ bps/Hz\cite{ding2024flexible}.  The adopted benchmark is a conventional-antenna system assisted by NOMA, where the BS is located at the center of the square region and is equipped with $N$ conventional antennas. These antennas are positioned at coordinates $\psi_{n}^\mathrm{C}=(x_{n}^\mathrm{C}, 0, h), n \in \mathcal{N}$, and are arranged at half-wavelength intervals. Furthermore, the channel vector between the conventional fixed antennas and User $m$ can be expressed as:

\begin{align}
\mathbf{h}_{m}^\mathrm{C} = \left[ \frac{\sqrt{\eta} e^{-j \frac{2 \pi}{\lambda} \left| \boldsymbol \phi_m - \psi_{1}^\mathrm{C} \right|}}{\left| \boldsymbol \phi_m - \psi_{1}^\mathrm{C} \right|}, \cdots, \frac{\sqrt{\eta} e^{-j \frac{2 \pi}{\lambda} \left| \boldsymbol \phi_m - \psi_{N}^\mathrm{C} \right|}}{\left| \boldsymbol \phi_m - \psi_{N}^\mathrm{C} \right|} \right]^\mathrm{T}.\label{Con_antennas}
\end{align}
\begin{figure}[t]
\begin{center}
\includegraphics[width=2.8in, height=2.0in]{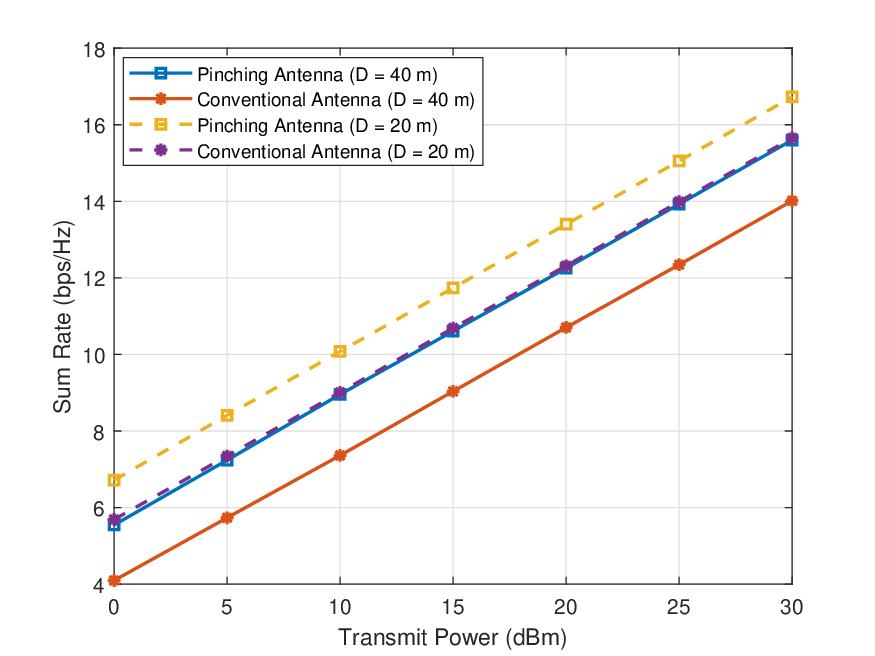}
\end{center}
\vspace*{-1.5em}
\caption{Achievable sum rate of pinching-antenna systems and conventional-antenna systems assisted by NOMA versus transmit power.}\label{g1}
\vspace{-0.7em}
\end{figure}

Fig. \ref{g1} illustrates the sum rate comparison between the NOMA-assisted pinching-antenna systems and the NOMA-assisted conventional-antenna systems under different transmit power levels. It is noteworthy that, for different deployment sizes, the sum rate of the pinching-antenna systems consistently is larger than that of the conventional-antenna systems. This is attributed to the ability of pinching antennas to flexibly reconfigure wireless channel conditions, creating the channel disparity for NOMA users and thereby achieving a higher sum rate compared to conventional fixed antennas. Furthermore, it can be seen from Fig. \ref{g1}, both the pinching-antenna systems and the conventional-antenna systems demonstrate improved performance in smaller deployment areas. This is due to the relative reduction in large-scale path loss between the BS and the users as the service area decreases, which enhances the overall system performance.

\begin{figure}[t]
\begin{center}
\includegraphics[width=2.8in, height=2.0in]{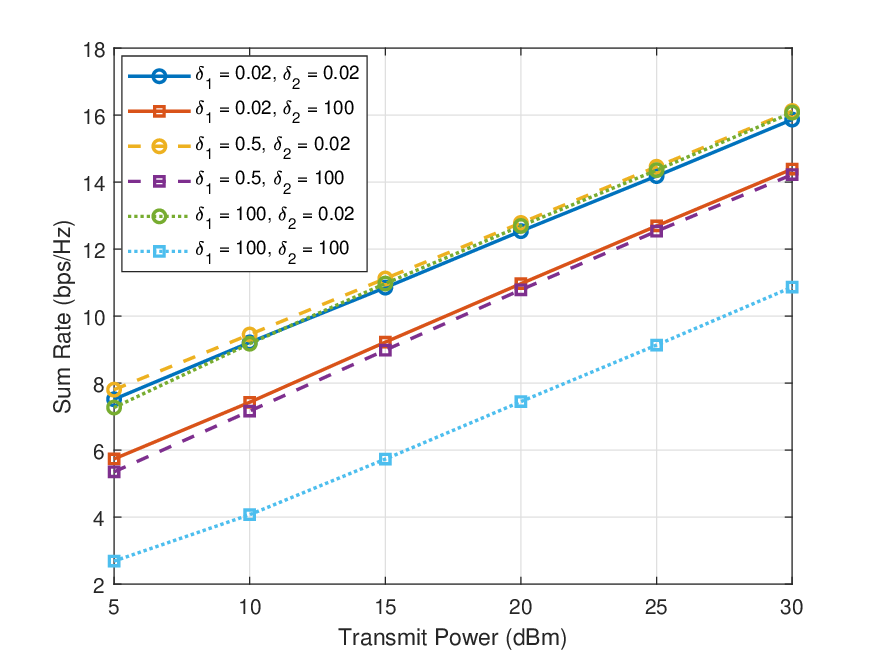}
\end{center}
\vspace*{-1.5em}
\caption{The effect of different phase alignment accuracies $\delta_1$ and $\delta_2$ on the achievable sum rate of NOMA-assisted pinching-antenna systems.}\label{g2}
\vspace{-0.7em}
\end{figure}

Fig. \ref{g2} illustrates the achievable sum rate of NOMA-assisted pinching-antenna systems under different phase alignment  accuracies, i.e., $\delta_1$ and $\delta_2$, highlighting the importance of fine-tuning antenna positions. It can be observed from Fig. \ref{g2} that improving the phase alignment accuracy of User $2$, i.e., from $\delta_2 = 100$ to $\delta_2 = 0.02$, significantly enhances the sum rate. This is because User $2$, as a strong user, plays a more critical role in sum rate optimization due to its favorable channel conditions. In addition, it is interesting to observe that when $\delta_1 = 0.5$ and $\delta_2 = 0.02$, the system achieves the best performance, rather than $\delta_1 = 0.2$ and $\delta_2 = 0.02$, suggesting that enforcing high-precision phase alignment for both users does not necessarily yield the results. This is because achieving high-precision  phase alignment for both users may require significant adjustments to the positions of the pinching antennas, which could negatively affect the large-scale fading of the users and ultimately degrade the overall system performance.

\begin{figure}[!htp]
\begin{center}
\includegraphics[width=2.8in, height=2.0in]{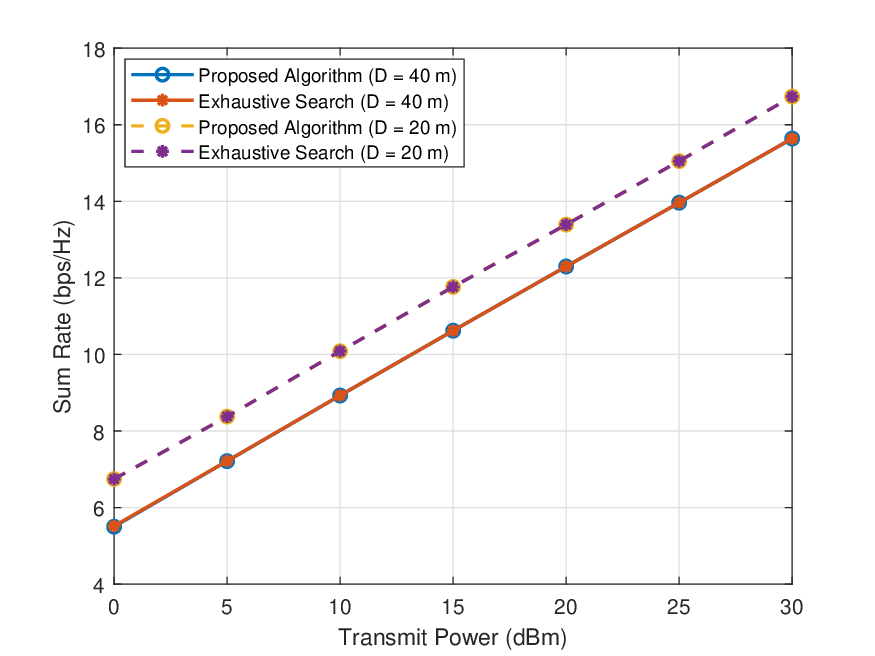}
\end{center}
\vspace*{-1.5em}
\caption{Comparison of the proposed algorithm with the exhaustive search method.}\label{g3}
\vspace{-0.7em}
\end{figure}

Fig. \ref{g3} illustrates the sum rate comparison between the proposed bisection-based algorithm and the exhaustive search method in NOMA-assisted pinching-antenna systems. Specifically, the exhaustive search method explores all possible positions of the pinching antennas on the waveguide with a step size of $\lambda/10$, under the condition that the physical spacing between the pinching antennas is at least half-wavelength. As seen in Fig. \ref{g3}, the sum rate of the proposed bisection-based iterative algorithm is almost identical to that of the exhaustive search method under different  deployment sizes. This demonstrates the feasibility and efficiency of the proposed algorithm, which first optimizes large-scale path loss to create optimal channel disparity for NOMA users, and then fine-tunes the antenna positions to achieve high-precision phase alignment, maximizing the sum rate.

\section{conclusion}
This letter investigated a downlink NOMA-assisted pinching-antenna system, aiming to maximize the sum rate by optimizing the power allocation coefficients for users and the positions of pinching antennas. To address this non-convex problem, we first derived a closed-form solution for the power allocation coefficients in relation to the pinching antenna positions. Subsequently, we proposed a low-complexity iterative algorithm based on the bisection method to determine the optimal deployment positions of the pinching antennas. Simulation results demonstrate that the NOMA-assisted pinching-antenna systems significantly outperforms the NOMA-assisted conventional-antenna systems in terms of sum rate. 

\bibliographystyle{IEEEtran}
\bibliography{NOMA_PASS}

\end{document}